\documentclass[structabstract]{aa}  
\usepackage{graphicx}
\usepackage{txfonts}
\usepackage{natbib}
\bibpunct{(}{)}{;}{a}{}{,} 
\usepackage{epsfig, color}
\usepackage[tight,hang]{subfigure}
\usepackage{isotope}
\usepackage{graphicx}
\usepackage{wasysym}
\usepackage{color}
\usepackage{units}

\newcommand{\msun}{\ensuremath{\mathrm{M}_\odot}}

\begin{document}
\title{A Subgrid-scale Model for Deflagration-to-Detonation Transitions in
  Type Ia Supernova Explosion Simulations}

\subtitle{Numerical implementation}

\author{F.~Ciaraldi-Schoolmann\inst{1} \and I.~R.~Seitenzahl\inst{1,2} \and F.~K.~R\"opke\inst{2}}

\titlerunning{A Subgrid-scale Model for DDT in SN~Ia Simulations}

\institute{Max-Planck-Institut f\"ur Astrophysik, Karl-Schwarzschild-Stra{\ss}e 1, D-85748 Garching, Germany \\
  \and
  Institut f\"ur Theoretische Physik und Astrophysik,
  Universit\"at W\"urzburg, Campus Hubland Nord,
  Emil-Fischer-Str. 31,\\
  D-97074 W\"urzburg, Germany\\ 
 \email{friedrich.roepke@astro.uni-wuerzburg.de}}

\date{Received xxxx xx, xxxx / accepted xxxx xx, xxxx}

 
\abstract
{A promising model for normal Type Ia supernova (SN Ia) explosions
  are delayed detonations of Chandrasekhar-mass white dwarfs, in which
  the burning starts out as a subsonic deflagration and turns at a
  later phase of the explosion into a supersonic detonation.  The
  mechanism of the underlying deflagration-to-detonation transition (DDT)
  is unknown in detail, but necessary conditions have been determined
  recently.  
  The region of detonation initiation cannot be
  spatially resolved in multi-dimensional full-star simulations of the explosion.}
{We develop a subgrid-scale (SGS) model for DDTs in thermonuclear
  supernova simulations that is consistent with the currently known
  constraints.}
{The probability for a DDT to occur is calculated from the distribution
  of turbulent velocities measured on the grid scale in the vicinity
  of the flame and the fractal flame surface area that satisfies
  further physical constraints, such as fuel fraction and fuel
  density.}
{The implementation of our DDT criterion provides a solid basis for
  simulations of thermonuclear supernova explosions in the delayed
  detonation scenario. It accounts for the currently known necessary
  conditions for the transition and avoids
  the inclusion of resolution-dependent quantities in the
  model. The functionality of our DDT criterion is demonstrated on
  the example of one three-dimensional thermonuclear supernova explosion simulation.}
{}


\maketitle
%

\section{Introduction}
\label{Sect.1}
In the Chandrasekhar-mass model for SNe~Ia, a thermonuclear burning
front (flame) ignites near the center of a white dwarf star when its
mass approaches the Chandrasekhar-limit \citep[see][for a review on
SNe~Ia models]{hillebrandt2000a}. In principle, there are two possible
modes for this flame to burn through the degenerate material: a
supersonic \emph{detonation} and a subsonic \emph{deflagration}. The
result of the thermonuclear burning process has to be consistent with
the main observational features, in particular the observed range in
brightness. The origin for the diversity in brightness of SNe~Ia are
primarily differences in the radioactive \isotope[56]{Ni} produced in
the explosion \citep{truran1967a,colgate1969a}. According to studies
of \citet{contardo2000a}, \citet{stritzinger2006a}, and
\citet{mazzali2007a}, any valid model for normal SN~Ia explosions
should cover a range in the \isotope[56]{Ni} production of ${\sim}0.4$
to $1.0\ \msun$.

Numerical simulations show that prompt detonations lead to strong
explosions that produce almost exclusively iron group elements
\citep{arnett1971a}, which is inconsistent with observed spectra. In
contrast, pure deflagrations produce not enough iron group elements
and release too little energy to explain the bulk of normal SNe~Ia
\citep{khokhlov2000a,gamezo2003a,roepke2007c}.  Moreover,
\citet{kozma2005a} argue that unburned material left behind by the
deflagration near the center of the star leaves imprints in nebular
spectra that are not observed in normal SNe~Ia.  These problems are
cured if a detonation triggers sometime during the late deflagration
phase.  In this \emph{delayed detonation scenario}
\citep{khokhlov1991a}, the detonation stage leads to a more complete
burning of the white dwarf, resulting in an explosion strength and a
chemical structure of the ejecta that is more consistent with the
observed characteristics of SNe~Ia \citep[e.g.][]{gamezo2005a,
  golombek2005a, roepke2007b, mazzali2007a, kasen2009a, roepke2012a,
  seitenzahl2011a,seitenzahl2013a}.

Whether or not a transition of the flame from a subsonic deflagration
to a supersonic detonation is possible in SNe~Ia has remained an open
question since \citet{blinnikov1986a} first alluded to such a
possibility.  To understand deflagration-to-detonation transitions
(DDTs) in general, the microphysical nature of turbulently mixed
flames has to be analyzed. Extensive studies in this field were
carried out by \citet{lisewski2000b}, \citet{woosley2007a},
\citet{aspden2008a}, and \citet{woosley2009a}.  Although these studies
do not provide stringent evidence for DDTs in SNe~Ia, necessary
conditions for such transitions can be derived from them.  In
particular, their analyses show that strong turbulence must interact
with the flame during later stages of the explosion in order to
facilitate a DDT.  This raises the question of whether sufficiently
high turbulent velocity fluctuations still occur when the deflagration
is close to extinction due to the expansion of the star.  The
Rayleigh-Taylor instability becomes weaker in the later expansion
phase, hence this expansion will ultimately freeze out all turbulent
motions \citep{khokhlov1995a}.  \citet{roepke2007d} showed that high
turbulent velocities, although rare, are indeed still found in late
stages of three-dimensional simulations of the deflagration phase.

This indicates that the \emph{macroscopic conditions} for a DDT are
met, but it is clear that evidence for DDTs requires to resolve the
microscopic mechanism of this transition as well. The length scales on
which this process takes place, however, are too small to be resolved
in multi-dimensional full-star simulations of the explosion.
Therefore, large-scale simulations of the delayed detonation scenario
have to invoke some kind of model for DDTs. A simple parameterization
is to prescribe a certain fuel density ahead of the flame at which the
DDT is triggered \citep{khokhlov1997a, hoeflich1998a, gamezo2005a,
  townsley2009a, jackson2010a}. This, however, does not account for
the important role that turbulence plays in the DDT mechanism. An
alternative is to trigger the DDT at patches of the burning front
where turbulent eddies first penetrate the internal flame structure
\citep{golombek2005a, roepke2007b}. The onset of this so-called
\emph{distributed burning regime} \citep[e.g.][]{peters2000a} is
necessary \citep{niemeyer1997b}, but still not sufficient for a
DDT. \citet{woosley2009a} argue that, in addition to entering the
distributed burning regime, particularly high velocity fluctuations
are required. In a very simple way this constraint has been
implemented in a series of two-dimensional delayed detonation
simulations \citep{kasen2009a}. Here, we present a subgrid-scale (SGS)
model of DDTs for full-star simulations of the delayed detonation
scenario. In particular, we aim at consistency with the microphysical
mechanism of this process, as far as known, and independence of the
numerical resolution in the simulation. Due to the stochastic nature
of turbulence, a SGS model for DDTs cannot provide any proof for a DDT
to occur, but it can evaluate a probability for this transition under
certain assumptions.

This paper is organized as follows. In Section~\ref{Sect.2} we outline
the constraints on DDTs in SNe~Ia according to current knowledge. The
implementation of the DDT-SGS model in the hydrodynamic code is
described in Section~\ref{Sect.3}. The resolution dependence of this
model is tested in Section~\ref{Sect.4}. Section~\ref{Sect.5} gives a
summary and an outlook for further applications.

\section{Constraints on DDT\lowercase{s} in
  SN\lowercase{e}~I\lowercase{a}}
\label{Sect.2}
Which physical mechanism causes a DDT in unconfined media (as required
in the supernova case) remains uncertain, but several possibilities
have been suggested.  One proposed
mechanism for the initiation of a detonation relies on the dissipation
and the consequential conversion of turbulent energy into internal
energy on the Kolmogorov length scale \citep{woosley2007a}. Here, it is
assumed that the rate of dissipating turbulent energy is high enough
that the temperature of a region of fuel reaches the ignition
point. Provided that a sufficient amount of fuel is available (the
ignition region is large enough) a detonation may be formed. Another
mechanism recently proposed by \citet{charignon2013a} is based on the
amplification of acoustic waves in the steep outer density gradient of
the white dwarf. This would trigger the detonation wave far away from
the deflagration front.
In our
work, however, we assume that the deflagration flame itself produces
conditions suitable for a DDT and follow the concept of the \emph{Zel'dovich Gradient
  Mechanism} \citep{zeldovich1970a}, even though it has been suggested
that the formation of a
preconditioned hot spot may not a neccessary prerequisite \citep{poludnenko2011a, kushnir2012a}. In the Gradient Mechanism,
it is assumed that a spontaneous ignition of the fuel in a region with
a shallow spatial gradient of induction times leads to a supersonic
reaction wave and the build-up of a shock. If the phase velocity of
the reaction wave approaches the Chapman-Jouguet velocity it may
transition into a detonation. The Gradient Mechanism has been applied
to SNe~Ia first by \citet{blinnikov1986a, blinnikov1987a} and has been
further investigated by \citet{khokhlov1991a, khokhlov1991c},
\citet{khokhlov1997a} and \citet{niemeyer1997b}. The most important
result of their analyses is that DDTs in SNe~Ia can only occur if
turbulence approaches an intensity that causes strong mixing of cold
fuel and hot burned material. A microphysical study of
\citet{lisewski2000b} revealed that the required turbulent velocity
fluctuations $v'_\mathrm{crit}$ must be higher than $\unit[10^8]{cm\,
  s^{-1}}$. By analyzing \emph{some} time steps of a pure deflagration
model, \citet{roepke2007d} found a non-vanishing probability of
finding such high velocity fluctuations at the flame. Hence, the
probability of finding sufficiently high velocity fluctuations in the
\emph{entire} late deflagration phase may reach high values.

The occurrence of high turbulent velocity fluctuations is attributed
to intermittency in the turbulent motions. Weak intermittency in
burned regions in the exploding white dwarf was found by
\citet{schmidt2010a} by calculating and fitting characteristic scaling
exponents of the turbulent velocity field. These exponents were
obtained from the computation of high-order velocity structure
functions \citep{ciaraldi2009a}, using the data of a highly resolved
numerical simulation, the deflagration model of
\citet{roepke2007c}. The high velocity fluctuations that
\citet{roepke2007d} found in the same model indicate that
intermittency at the flame is significantly stronger than in burned
regions. However, due to the challenges of performing a detailed
analysis of intermittency at a highly wrinkled and folded flame front
in full-star simulations, some uncertainties in the origin of these
high velocity fluctuations remain.

That high velocity fluctuations occur \emph{somewhere} at the flame is
necessary, but not sufficient for a DDT.  It is important that these
fluctuations are located within a certain amount of fuel of the
turbulently mixed regions.  The minimum amount of fuel
$X_\mathrm{fuel}^\mathrm{DDT}$ required for ignition and creation of a
self-propagating detonation wave depends on various quantities, such
as the fuel density, the chemical composition, as well as the fuel
temperature \citep[see][]{arnett1994b, khokhlov1997a,
  seitenzahl2009b}. Due to these dependencies, one cannot specify a
general, constant value for $X_\mathrm{fuel}^\mathrm{DDT}$ \citep[but
see][and tables therein]{seitenzahl2009b}.

\citet{niemeyer1997b} point out that a necessary constraint for a DDT
is the burning in the distributed burning regime. 
In the distributed burning regime,
turbulent eddies are able to penetrate the internal flame
structure. Under this condition, the nuclear burning time scale
$\tau_\mathrm{nuc}$ becomes independent of heat conduction processes
and is exclusively given by the dynamics of turbulent eddies. The
reason is that these eddies reach the fuel faster than the flame
itself and mix it during the turnover into the reaction zone. The eddy
turnover time is given by
\begin{equation}
  \label{tau}
  \tau_\mathrm{eddy}(\ell) = \ell/v'(\ell) ,
\end{equation}
where $\ell$ is the typical length scale of a turbulent eddy and
$v'(\ell)$ the velocity fluctuation on that scale.

\citet{woosley2007a} point out that for a successful DDT, 
  the carbon and the oxygen flame have to be sufficiently separated
  spatially. They argue that this is expected to be the case for fuel
  densities below $\sim$$3\times 10^7 \,\mathrm{g}\,\mathrm{cm}^3$,
  which covers the density regime we consider here.

The distributed burning regime for the canonical composition of equal
mass $^{12}$C and $^{16}$O is reached when the fuel density
$\rho_\mathrm{fuel}$ at the flame has declined below ${\sim}\unit[3
\times 10^7]{g\ cm^{-3}}$ \citep{niemeyer1997b}. Recent studies of
\citet{woosley2007a} and \citet{woosley2009a} suggest that there are
further constraints on triggering detonations. Within the distributed
burning regime, it is necessary that the balance between turbulent
mixing and nuclear burning becomes disturbed, which is the case for
$D_\mathrm{T} = \tau_\mathrm{eddy}(L) / \tau_\mathrm{nuc} \gtrsim 1$,
where $D_\mathrm{T}$ is the turbulent Damk\"ohler number and $L$ the
turbulent integral scale. During the burning in this so-called
\emph{stirred flame regime} \citep{kerstein2001a}, the flame becomes
significantly broadened until at $D_\mathrm{T} \sim 1$ the flame width
$\delta$ approaches $L$ which is approximately $\unit[10^6]{cm}$
\citep[e.g.][]{woosley2007a}. With turbulent intensities typically
expected for deflagrations in SNe~Ia, the density at which this
condition is expected to be met is $0.5 \lesssim \rho_\mathrm{fuel} /
(10^7 \mathrm{g\ cm^{-3}}) \lesssim 1.5$ \citep{woosley2007a}.

Finally, a DDT region which meets the described constraints concerning
$v'_\mathrm{crit}$, $X_\mathrm{fuel}^\mathrm{DDT}$ and
$\rho_\mathrm{fuel}$ has to exceed a critical spatial scale
$\ell_\mathrm{crit}$, which is of the order of $10^6$ cm
\citep[e.g.][]{khokhlov1997a, dursi2006a, seitenzahl2009b} and hence
comparable to the integral scale $L$. The time scale of mixing the
fuel and ash in this region can be estimated with
Eq.~(\ref{tau}). Assuming that both fuel and ash elements can be
carried by a turbulent eddy of size $\ell_\mathrm{crit}$ over the
distance $\ell_\mathrm{crit}$ in a half eddy turnover time, it takes
\begin{equation}
  \label{tau2}
  \tau_\mathrm{eddy_{1/2}}(\ell_\mathrm{crit}) =
  \tau_\mathrm{eddy}(\ell_\mathrm{crit})/2= \ell_\mathrm{crit}/2
  v'(\ell_\mathrm{crit})
\end{equation}
to mix the components. While $v'$ is well-determined in our model,
$\ell_\mathrm{crit}$ is uncertain because of the unresolved shape of the
temperature gradient
  \citep{seitenzahl2009b}. 
Using $v'(\ell_\mathrm{crit}) = \unit[10^8]{cm\,
  s^{-1}}$ \citep{lisewski2000b}, we find
$\tau_\mathrm{eddy_{1/2}}(\ell_\mathrm{crit}) = \unit[5 \times
10^{-3}]{s}$ and we adopt this typical, fixed value in our model. A region fulfilling all DDT criteria described above
must exist for at least this amount of time such that a DDT may occur.

\section{Formulation of a subgrid-scale model for DDT\lowercase{s}}
\label{Sect.3}

\subsection{Three-dimensional full-star simulations}
\label{Sect.3.1}
The hydrodynamics code that is used to carry out the simulations of
this study is based on the {\sc{PROMETHEUS}} code \citep{fryxell1989a}
that implements the Piecewise Parabolic Methods (PPM) of
\citet{colella1984a} to solve the reactive Euler equations in a finite
volume approach. The thermonuclear combustion waves are modeled as
sharp discontinuities between fuel and ash and are numerically
represented with a level set technique following
\citet{reinecke1999a}.  Our implementation follows some basic concepts
of large eddy simulations, in which the largest turbulent structures
and motions are resolved on the grid scale or above. Turbulence on
unresolved scales is calculated with a SGS turbulence model
\citep{schmidt2006b,schmidt2006c}.  In our simulations, we use a
comoving grid technique \citep{roepke2005c,roepke2006a}. We discretize
our set of model equations on two nested computational grids for which
the grid spacing is continuously enlarged to capture the explosion.
While an outer inhomogeneous grid follows the overall expansion of the
white dwarf, the deflagration flame is tracked with an inner
homogeneous Cartesian grid.

For the initial composition of the white dwarf, we choose a $^{12}$C
and $^{16}$O mixture in equal amounts by mass and set the electron
  fraction to $Y_e = 0.49886$, 
corresponding to solar metallicity. The white dwarf is
assumed to be cold ($T=5 \times 10^5 \, \mathrm{K}$).  We use an initial
central density of $\unit[2.9\times 10^9]{g\ cm^{-3}}$.  The initial
flame configuration from which the deflagration front evolves equals
the setup described in \citet{roepke2007c} with 1600 spherical
  kernels of radius $2.6\,\mathrm{km}$ distributed within a sphere of
  $180\, \mathrm{km}$ around the center of the white dwarf.  In our full-star
simulations, the DDT regions are not resolved, since $\Delta(t) >
\ell_\mathrm{crit}$ for all times, where $\Delta(t)$ is the
time-dependent resolution of the inner comoving grid.  Therefore, we
employ a SGS model for DDTs, which models the DDT relevant quantities
on unresolved scales.

\subsection{Determination of the flame surface area}
\label{Sect.3.2}

As described in Section~\ref{Sect.2}, we have to determine the area of
the flame where the values of $X_\mathrm{fuel}^\mathrm{DDT}$,
$\rho_\mathrm{fuel}$ and $v'_\mathrm{crit}$ are appropriate for a
DDT. Here we face the problem that the discontinuity approach of the
flame generally prevents us from determining the physical conditions
at the flame precisely.  Below we show how to obtain an approximation
for the physical conditions at the flame front, by considering only
grid cells that are approximately split into two equal parts by the
flame (resp. the level set).

We define $X_\mathrm{fuel}$ as the mass fraction of unburned material
in a grid cell.  For the later analysis, we are interested in the
  quantities \emph{at the flame}. These are difficult to measure since
  the flame is numerically represented as a discontinuity and the
  computational cells intersected by it contain a mixture of fuel and
  ash. We therefore consider only cells with $1/3 \leq X_\mathrm{fuel}
  \leq 2/3$.
This way we ensure that the flame separates the grid cell into roughly
equal size parts of fuel and ash, and the thermodynamic values at the
cell center should reasonably approximate the real values at the turbulent
flame, instead of being dominated by fuel or ash material.
We emphasize that the \emph{numerical} quantity $X_\mathrm{fuel}$ is
not directly equivalent to the required \emph{physical} amount of fuel
$X_\mathrm{fuel}^\mathrm{DDT}$ for triggering a DDT. $X_\mathrm{fuel}^\mathrm{DDT}$ cannot accurately be determined
on scales $\ell_\mathrm{crit} < \Delta(t)$ and we cannot evaluate
precisely whether the required amount of fuel for a DDT is available.

As described in Section~\ref{Sect.2}, we further have to ensure that
the flame resides in the distributed burning regime and additionally
obeys the constraints described by \citet{woosley2007a}. Therefore we
additionally limit our analysis to grid cells in the density range of
$0.5 \lesssim \rho_\mathrm{fuel} / (10^7 \mathrm{g\ cm^{-3}}) \lesssim
1.5$.

We define the number of all grid cells at the flame at a given time
$t$ as $N_\mathrm{flame}(t)$ and the cells which additionally meet the
constraints concerning $X_\mathrm{fuel}$ and $\rho_\mathrm{fuel}$ as
$N_\mathrm{flame}^{*}(t)$. In the same context we define the flame
surface area as $A_\mathrm{flame}(t)$ and the part which meets the
mentioned constraints as $A_\mathrm{flame}^{*}(t)$, respectively. To
determine $A_\mathrm{flame}(t)$ we assume that due to the nature of
turbulence the flame is similar to a fractal object with fractal
dimension $D$ \citep[see][]{kerstein1988a,kerstein1991b,niemeyer_phd,blinnikov1996a}. 
We note that compared to an ideal fractal, the wrinkles and curvatures
of the flame are not sustained on very small scales.

In our model, the DDT occurs shortly after entering the distributed regime. Stricly speaking, the description of the flame as a fractal was established for the flamelet regime only. However, for the specific case we consider here, the flame neither fills the entire star nor a large fraction of its volume.
Instead, seen from the large scales resolved in our simulations, the 
burning is still confined to a narrow sheet, to which we apply our fractal description. The same line of argument was used by \citet{schmidt2007a} to justify a level-set based flame model beyond the flamelet regime.
Therefore, for our large-scale simulations, a fractal approach is an acceptable description of the flame for all physical scales directly relevant to our DDT model.

If turbulence is driven by the Rayleigh-Taylor instability, \mbox{$D =
  2.5$}, whereas for Kolmogorov turbulence without intermittency, a
value of \mbox{$D = 2.33$} is expected \citep[e.g.][and references
therein]{kerstein1988a, sreenivasan1991a, kerstein1991b,
  niemeyer_phd}.  For intermittent turbulence, it is argued that
\mbox{$D = 2.36$} \citep[e.g.][]{halsey1986a, sreenivasan1991a}.

The level set method offers us the opportunity to relate the
quantities $\Delta(t)$ and $N_\mathrm{flame}(t)$ to
$A_\mathrm{flame}(t)$. Since for every numerical resolution the flame
propagates like a thin interface through the grid cells, we assume
that the flame surface behaves self-similar and is
resolution-independent on all considered length scales. We therefore
determine the self-similarity dimension defined by

\begin{equation}
  \label{D}
  D = \frac{\log N}{\log\epsilon}
\end{equation}
where $N$ is the number of self-similar pieces and $\epsilon$ the
reduction (or zoom) factor. For our purposes we need the number of
grid cells $N_\mathrm{flame_1}$ and $N_\mathrm{flame_2}$ from two
simulations with different resolutions $\Delta_1(t)$ and $\Delta_2(t)$
of the same initial white dwarf model. Then $D$ is given by
\begin{equation}
  \label{eq5}
  D = \frac{\log[{N_\mathrm{flame_2}(t)/N_\mathrm{flame_1}(t)}]}{\log[{\Delta_1(t)/\Delta_2(t)}]}. 
\end{equation}
From here it follows
\begin{equation}
  \label{eq4}
  N_\mathrm{flame_1}(t) \Delta_1(t)^D = N_\mathrm{flame_2}(t) \Delta_2(t)^D 
\end{equation}
and since $A_\mathrm{flame}(t)$ should be equal for both simulations,
we identify
\begin{equation}
  \label{eq3}
  A_\mathrm{flame}(t)\approx N_\mathrm{flame}(t) \Delta(t)^D
\end{equation}
as the flame surface area. Once $D$ is determined we evaluate
$A_\mathrm{flame}^{*}(t)$ with Eq.~\ref{eq3} by using
$N_\mathrm{flame}^{*}(t)$ instead of $N_\mathrm{flame}(t)$. Since
$N_\mathrm{flame}^{*}(t)\ll N_\mathrm{flame}(t)$ there are not enough
data to derive a reliable value of $D$ for $A_\mathrm{flame}^{*}(t)$
directly. The calculation of $D$ is performed together with a
resolution test in Section~\ref{Sect.4.1}.

\subsection{The probability density function of turbulent velocity
  fluctuations}
\label{Sect.3.3}

The turbulent velocity fluctuations $v'(\ell)$ are determined by the
SGS model of \citet{schmidt2006b,schmidt2006c}. This model has already
been applied to a simulation of a pure deflagration in a
Chandrasekhar-mass WD \citep[e.g.][]{roepke2007c}, and turbulence
properties of this model were analyzed in
\citet{ciaraldi2009a}. However, it has not been explicitly tested yet
whether the SGS model can properly reproduce the rare high velocity
fluctuations at the flame required for a DDT. In this section we
perform some test calculations in order to evaluate whether the SGS
model can be used for the construction of a DDT model.

\subsubsection{Testing the SGS model in reproducing the high velocity
  fluctuations}
\label{Sect.3.3.1}

To judge whether the SGS model is capable of modeling the high
velocity fluctuations at the flame correctly, we first have to find
out how often these fluctuations occur. A commonly used statistical
method is the calculation of a probability density function (PDF) of
$v'(\ell)$. By definition, a PDF constitutes a continuous distribution
function, but in our case only discrete data are available.  However,
by sorting and sampling the data into bins, we can construct a
histogram of $v'(\ell)$.  Fitting this histograms with an appropriate
fit function then gives us an approximated PDF of $v'(\ell)$.  This
procedure has already been performed by \citet{roepke2007d}. The
result shows clearly a slow decline of the histogram toward higher
velocity fluctuations, indicating a non-vanishing probability of
finding sufficiently high velocity fluctuations for a DDT.  However,
an open question is whether the slow decline seen in the histogram is
of physical origin, or whether it is an artifact of turbulence- or
flame-modeling. To investigate this, we developed an algorithm that
derives the velocity fluctuations from the resolved velocity field of
the hydrodynamic flow. This allows us to compare the histogram that
contains the data of these resolved fluctuations with the histogram
that contains the values $v'(\ell)$ of the SGS model.

The resolved velocity field $\mathbf{v}(\mathbf{r})$ of the
hydrodynamic flow is a superposition of the turbulent velocity
fluctuations and the bulk expansion of the white dwarf, where the
latter contribution points in radial direction. We have to subtract
the bulk expansion from $\mathbf{v}(\mathbf{r})$ to obtain the pure
fluctuating part $\mathbf{v}_\mathrm{turb}(\mathbf{r})$. For details
on how the turbulent velocity fluctiations are calculated see
\citet{ciaraldi2009a}.  To compare $v_\mathrm{turb}(\mathbf{r})=
|\mathbf{v}_\mathrm{turb}(\mathbf{r})|$ with $v'(\ell)$, we have to
take into account that the SGS model returns a value on the scale
$\Delta(t)$ and that the quantity $v_\mathrm{turb}(\mathbf{r})$ has to
be considered on the same scale. We thus determine the average
absolute velocity differences $\overline{|v_\mathrm{turb}[\Delta
  (t)]|}$ of neighboring grid cells, which is given by
\begin{equation}
  \overline{|v_\mathrm{turb}[\Delta (t)]|} =
  \frac{1}{N}\sum^N_{i=1}|v_\mathrm{turb}(\mathbf{r})-v_\mathrm{turb_i}(\mathbf{r}
  + \mathbf{d})|
  \label{Eq5.05}
\end{equation}
where $v_\mathrm{turb}(\mathbf{r})$ is the velocity fluctuation in the
chosen grid cell and $v_\mathrm{turb_i}(\mathbf{r} + \mathbf{d})$ is
the velocity fluctuation in the $i$-th of the $N$ adjacent grid cells
(note that $|\mathbf{d}|=\Delta(t)$). The described procedure has been
performed with a Monte-Carlo based program for a total number of
randomly chosen $10^6$ different grid cells, where for a larger number
of cells, no change in the results was found. We then construct a
histogram of $\overline{|v_\mathrm{turb}[\Delta (t)]|}$.

\begin{figure*}
  \begin{center}
    \subfigure[Comparison of the histograms that contain the data at
    the flame of the resolved velocity fluctuations
    $\overline{|v_\mathrm{turb}(\ell)|}$ for the scales $\ell = \Delta
    (t)$ and \mbox{$\ell = 4 \Delta (t)$} and of the velocity fluctuations
    $ v' {[} \Delta (t) {]} $ of the turbulence SGS model.]
    {\includegraphics[angle=270,width=\columnwidth]{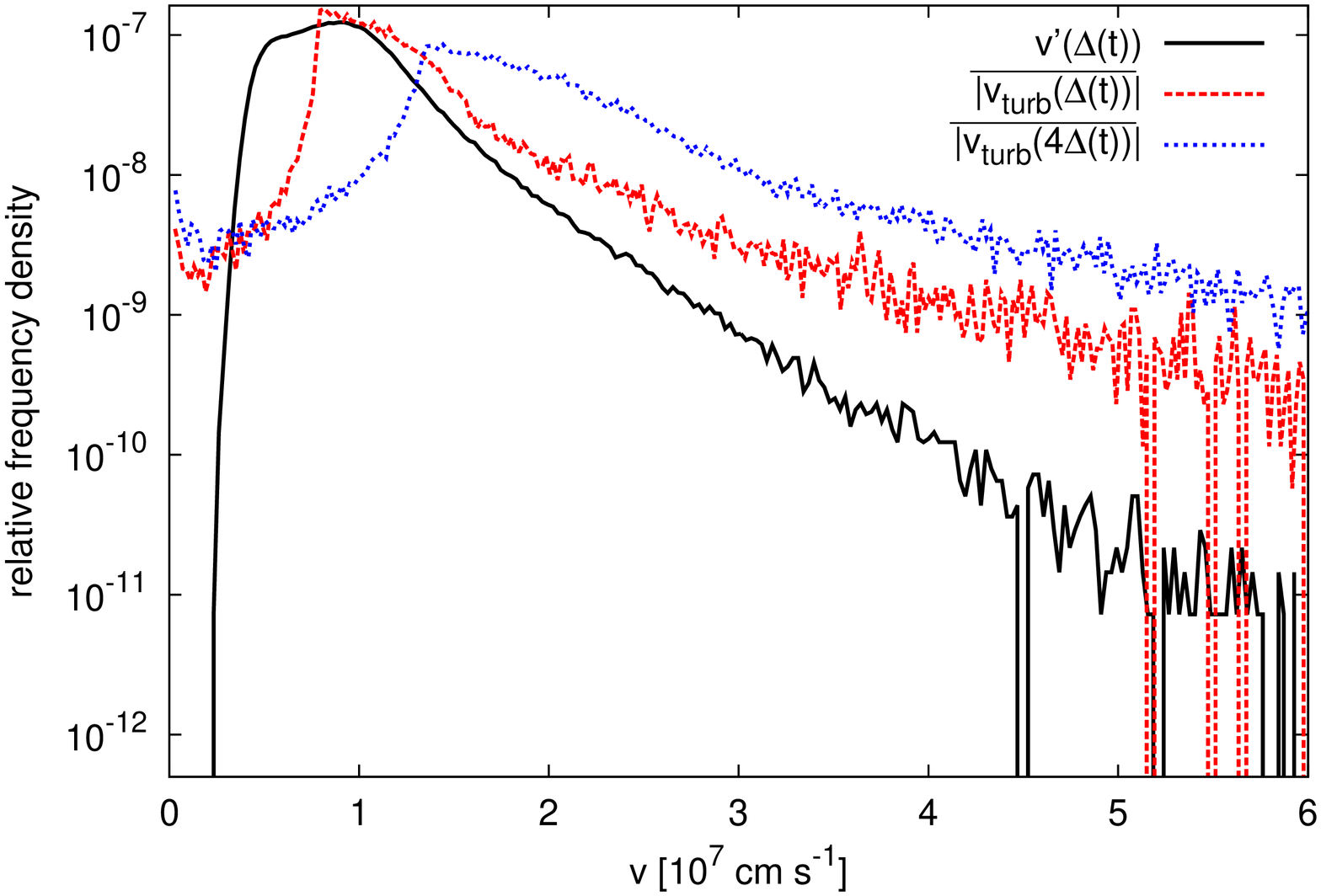}}
    \subfigure[Histogram of $v'(\ell_\mathrm{crit})$ using a rescaling
    factor of $\alpha = 1/3$ (Kolmogorov) and $\alpha = 1/2$
    (Rayleigh-Taylor instability) using $256^3$ grid cells.]
    {\includegraphics[angle=270,width=\columnwidth]{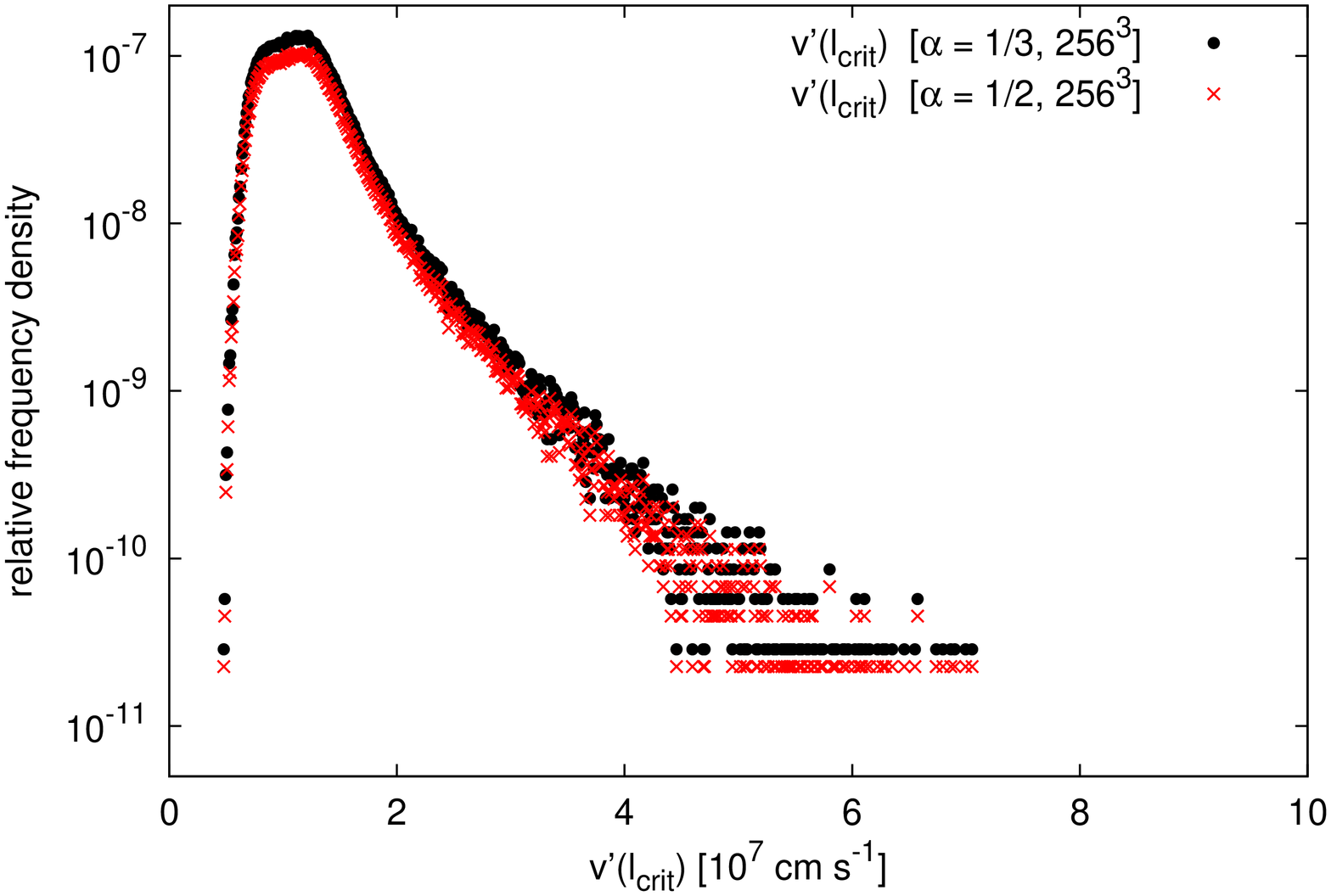}}
    \subfigure[Histogram of $v'(\ell_\mathrm{crit})$ using a rescaling
    factor of $\alpha = 1/3$ (Kolmogorov) and $\alpha = 1/2$
    (Rayleigh-Taylor instability) using $512^3$ grid cells.]
    {\includegraphics[angle=270,width=\columnwidth]{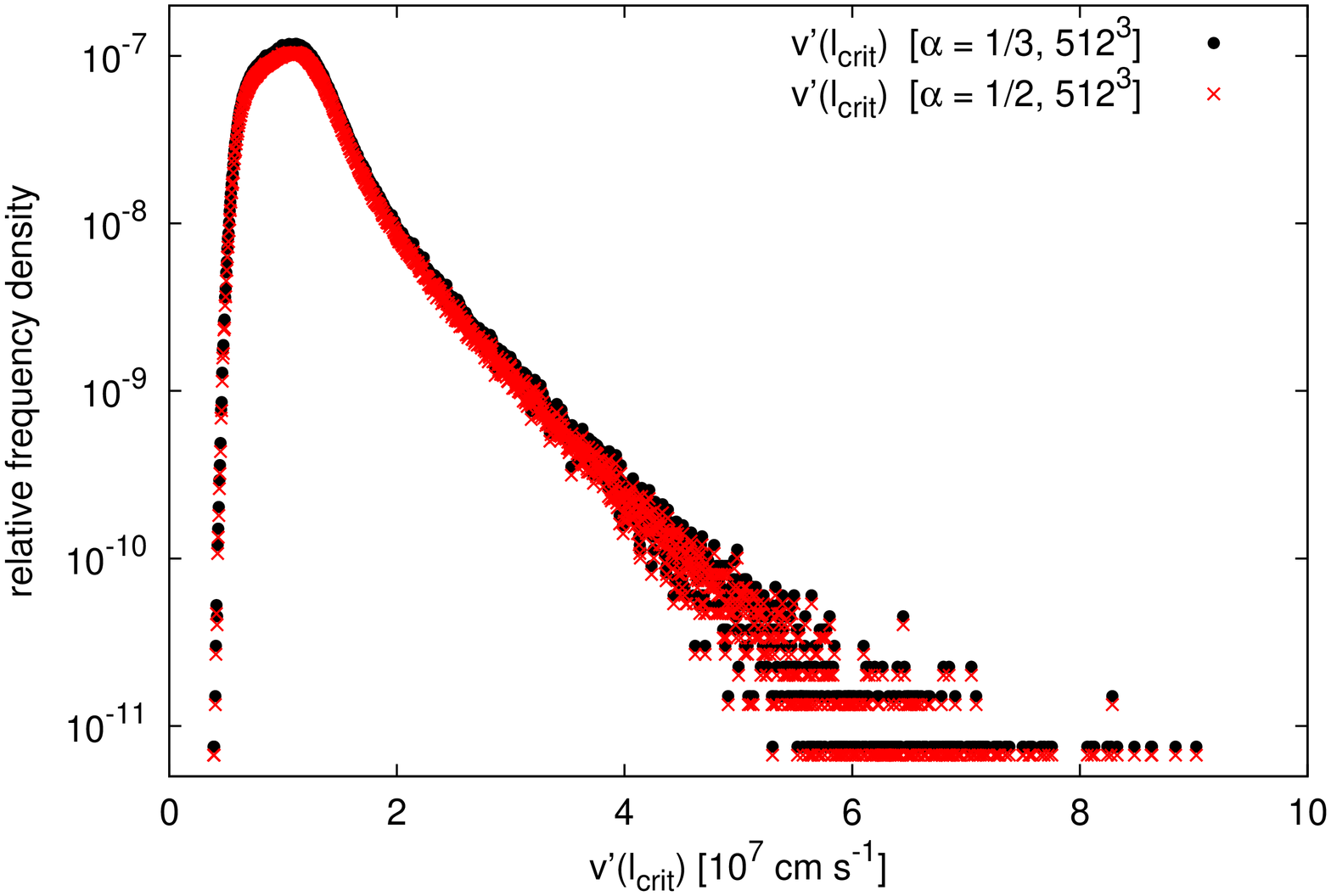}}
    \subfigure[Comparison of the histograms of
    $v'(\ell_\mathrm{crit})$ that contain the data at the flame front
    and in ash regions.]
    {\includegraphics[angle=270,width=\columnwidth]{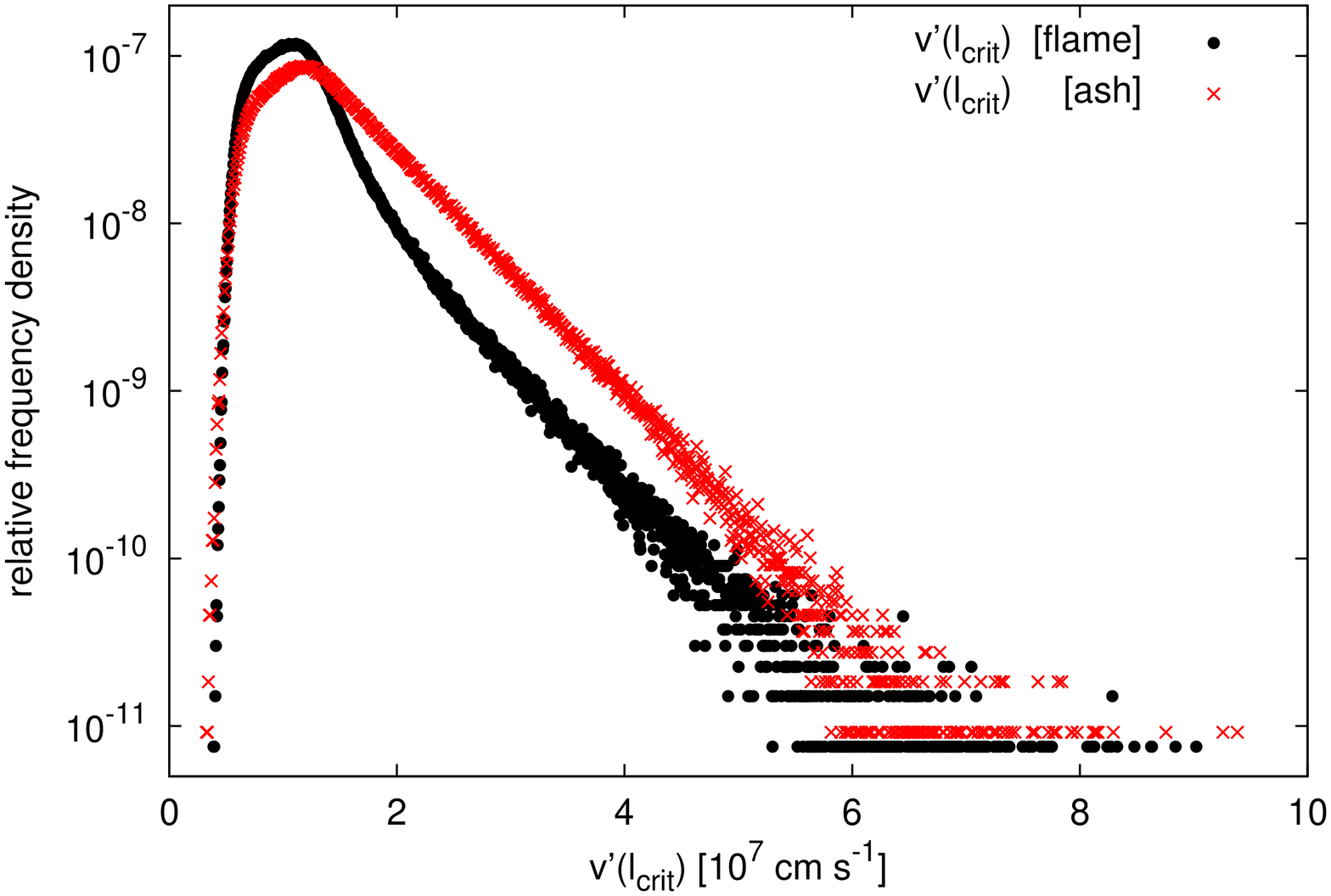}}
    \caption{Histograms of the turbulent velocity fluctuations at
      $t\approx 0.9\,\mathrm{s}$. (a) In
      the histograms that contain the data of velocity fluctuations
      from the hydrodynamic flow, a slow decline toward high velocity
      fluctuations is found for both length scales $\ell = \Delta (t)$
      and $\ell = 4 \Delta (t)$. This decline is similar to that of
      the histogram of $v'[\Delta(t)]$, revealing that the
      implemented turbulence model of
      \citet{schmidt2006b,schmidt2006c} calculates the high velocity
      fluctuations at the flame in a correct way. (b,c) A different
      rescaling of $v'(\ell_\mathrm{crit})$ using a rescaling factor
      of $\alpha = 1/3$ (Kolmogorov) and $\alpha = 1/2$
      (Rayleigh-Taylor instability) lead to some deviations,
      particularly for the lower resolved simulation with $256^3$ grid
      cells. (d) The different shapes of the histograms of
      $v'(\ell_\mathrm{crit})$ that contain the data in ash regions
      and at the flame front indicate that intermittency is stronger
      in the latter case \citep[see also][]{schmidt2010a}.}
    \label{fig:1}
  \end{center}
\end{figure*}

In Fig.~\ref{fig:1}(a) the histograms of
$\overline{|v_\mathrm{turb}[\Delta(t)]|}$ and $v'[\Delta(t)]$ that
contain the data in the vicinity of the flame are shown. The
simulation is based on a grid with $512^3$ cells and the histograms
shown are for $t\approx\unit[0.9]{s}$ as an illustrative example. This
instant corresponds to the late
deflagration phase, when turbulence is strong and affects the
structure and propagation of the flame significantly.  We see in both
histograms a slowing decline toward higher velocity fluctuations,
which shows that the decline in the histogram of $v'[\Delta(t)]$ is no
artifact of SGS turbulence model. Another possibility, however, is
that it is caused by our level-set based flame model and the
flame-flow coupling on the resolved scales. We therefore repeat the
analysis described above using a fixed length scale of $|\mathbf{d}|=4
\Delta(t)$. Even though the turbulence model calculates quantities on
the grid scale, in this case a rescaling of the velocity fluctuation
from $\Delta(t)$ to $4 \Delta(t)$ is not required for evaluating the
presence of the highest velocity fluctuations in the tail of the
histogram.  For $|\mathbf{d}|=4 \Delta(t)$ we impose the additional constraint
$X_\mathrm{fuel} \leq 0.5$ to avoid counting cells containing mainly
fuel far ahead of flame.  This result is also shown in
Fig.~\ref{fig:1}(a). We can identify again a slow decline toward high
velocity fluctuations similar to the histogram of
$\overline{|v_\mathrm{turb}[\Delta(t)]|}$, and hence also to that of
$v'[\Delta(t)]$. Thus, the slow decline seems to originate not only
from computational cells that are intersected by the flame but it
persists in a certain region away from it. This indicates that it is
not an artifact of the modeling but is rather due to intermittency in
the turbulent flow field near the flame.

\subsubsection{Rescaling of the velocity fluctuations}
\label{Sect.3.3.2}
Since our simulation code uses a comoving grid technique, we rescale
the value of $v'[\Delta(t)]$ to $v'(\ell_\mathrm{crit})$ with
$\ell_\mathrm{crit} = 10^6$ cm (see Section~\ref{Sect.2}). The
rescaled velocity fluctuations are given by
\begin{equation}
  \label{eq6}
  v'(\ell_\mathrm{crit}) = v'[\Delta(t)] [\ell_\mathrm{crit}/\Delta(t)]^\alpha
\end{equation}
where the scaling exponent $\alpha$ depends on the mechanism which
drives the turbulence. We assume incompressible and isotropic
Kolmogorov turbulence \citep{kolmogorov1941a}, where $\alpha = 1/3$.
We note, however, that \citet{ciaraldi2009a} found in burned regions a
transition of the turbulence driving mechanism at a certain length
scale \citep[see also][]{niemeyer1997b}. This length scale is of the
same order of magnitude as $\ell_\mathrm{crit}$ and it separates the
regime of small-scale isotropic Kolmogorov turbulence from
Rayleigh-Taylor instability driven anisotropic turbulence on large
scales. For the latter, $\alpha = 1/2$. These considerations take the
entire turbulent velocity field into account that has well-defined
statistical properties, but for a DDT only the strong turbulent
velocity fluctuations are important. Turbulence is most intense in
trailing patches of the Rayleigh-Taylor ``mushroom caps'', where
strong shear instabilities occur \citep[see][]{roepke2007d}. The
scaling properties of an intermittent velocity field for scales $\ell
\gtrsim \ell_\mathrm{crit}$ in such regions at the flame front are not
known. We can estimate the difference $f_\mathrm{diff}$ between the
scaling relations of a Kolmogorov- and Rayleigh-Taylor instability
driven turbulence. Using Eq.~\ref{eq6} we find
\begin{equation}
  \label{eq7}
  f_\mathrm{diff} = \frac{\left[\Delta(t)/\ell_\mathrm{crit}\right]^{1/2}}{\left[\Delta (t)/\ell_\mathrm{crit}\right]^{1/3}}=\left[\Delta(t)/\ell_\mathrm{crit}\right]^{1/6} .
\end{equation}
For highly resolved simulations, where
$\Delta(t)\approx\ell_\mathrm{crit}$, the difference is negligible.
We perform simulations with $256^3$ and $512^3$ grid cells and find
for the late deflagration phase where DDTs are expected
$\Delta(t)\approx\unit[4 \times 10^6]{cm}$ for the lower resolved
and $\Delta(t)\approx\unit[2 \times 10^6]{cm}$ for the higher
resolved simulation, leading to uncertainties of about $ 26\%$ and $12
\%$, respectively. To check to what extent these deviations affect the
rescaled values of the high velocity fluctuations, we compare the
histograms of $v'(\ell_\mathrm{crit})$ with both scaling exponents
$\alpha = 1/3$ and $\alpha = 1/2$. Since we implement a DDT model we
take now only grid cells into account that meet certain DDT
constraints, hence the data $N_\mathrm{flame}^{*}(t)$ is used for the
histogram construction. The result is shown in Fig.~\ref{fig:1}(b,c)
for the late deflagration phase at $t\approx\unit[0.9]{s}$.  The
agreement of both histograms is excellent, particularly in the high
resolution case.

We note that intermittency may slow down the decrease of the velocity
fluctuations towards smaller scales compared to the scaling given in
eq.~\ref{eq6}, or, if it dominates the scaling behavior, it may change
the trend completely. Our model would still be a good approximation in
the first case. Comparing the histograms in Fig.~\ref{fig:1}(a)
suggests that indeed the velocity fluctuations still decrease with
scale, but a more rigorous verification is not possible with our
simulations. While studying intermittency effects in ash regions
  is possible based on the computation of structure functions of the
  velocity field \citep{schmidt2010a, ciaraldi2009a}, for geometrical
  reasons such functions cannot easily be determined at the flame
  front itself.

\subsubsection{Fitting the data of the histogram}
\label{Sect.3.3.3}

To calculate the probability of finding sufficiently high velocity
fluctuations for a DDT, we apply a fit to the histogram of
$v'(\ell_\mathrm{crit})$ to obtain an approximated PDF \citep[see
also][]{roepke2007d}. Since for a DDT only the high velocity
fluctuations are of interest, we are justified in restricting our fit
to the right of the maximum of the histogram. The fit should further
be motivated by an appropriate distribution function that can explain
the intermittent behavior in turbulence at the
flame. \citet{schmidt2010a} used a log-normal distribution of an
intermittency model of \citet{kolmogorov1962a} and
\citet{oboukhov1962a} to fit characteristic scaling exponents that
where obtained from the computation of high order velocity correlation
functions. This detailed analysis revealed that the intermittency in
ash regions is weaker than predicted in the log normal model. In
contrast, \citet{roepke2007d} found that a log-normal fit fails to
reproduce the distribution of the high velocity fluctuations at the
flame, since it declines faster toward larger $v'(\ell_\mathrm{crit})$
than the velocity data of the histogram. This result suggests that
intermittency at the flame is fundamentally different than in ash
\citep[see also the discussion in][]{schmidt2010a}. In
Fig.~\ref{fig:1}(d) we show histograms of $v'(\ell_\mathrm{crit})$
that contain the data $N_\mathrm{flame}^{*}(t)$ and the data in ash
regions in the late deflagration phase at $t\approx 0.9 s$ (again,
this instant is chosen as an illustrative example here). The
simulation was run with $512^3$ grid cells. There is a significant
difference between the shapes of the PDFs. The slow decline of the
histogram that contains the data in ash regions appears almost linear
in the log-normal illustration, while the histogram that contains the
data in the vicinity of the flame has a significant positive curvature
after its maximum. This is further evidence that turbulence near the
flame has stronger intermittency than in ash regions.

As of yet there is no physically motivated model for explaining
intermittency at a deflagration front in white dwarfs. Consequently,
an empirical distribution function has to be used to fit the slow
decline of the histogram of $v'(\ell_\mathrm{crit})$ at the flame
front. Here we follow \citet{roepke2007d} and use an ansatz of the
form
\begin{equation}
  \label{eq9} f[v'(\ell_\mathrm{crit})]=\exp\lbrace a_1 [v'(\ell_\mathrm{crit})]^{a_2}+a_3\rbrace .
\end{equation}
This geometric function is able to fit the right part of the histogram
over a large range and $a_1$, $a_2$ and $a_3$ are the three fitting
paramters.  The probability \mbox{$P[v'(\ell_\mathrm{crit}) \geq
  v'_\mathrm{crit}](t)$} of finding velocity fluctuations of at least
$v'_\mathrm{crit}$ is given by
\begin{eqnarray}
  \label{eq10}
  P[v'(\ell_\mathrm{crit}) \geq v'_\mathrm{crit}](t)=\int^{\infty}_{v'_\mathrm{crit}}f[v'(\ell_\mathrm{crit})]~dv'(\ell_\mathrm{crit})\nonumber\\
=\frac{\exp(a_3)\Gamma(1/a_2,-a_1v^{a_2})}{a_2(-a_1)^{1/a_2}}
\end{eqnarray}
where $\Gamma$ is the upper incomplete gamma function.

We note that the DDT instant determined below is not really a special
point in the time evolution of the PDF. When a detonation is triggered
in the model, the parts of the deflagration flame that are directly
attached to the quickly spreading detonation front are excluded from
the determination of the PDF.

\subsection{The detonation area and the DDT criterion}
\label{Sect.3.4}

In Section~\ref{Sect.3.2} we defined $A_\mathrm{flame}^{*}(t)$ as the
part of the flame that meets the required conditions for a DDT
concerning the quantities $\rho_\mathrm{fuel}$ and
$X_\mathrm{fuel}$. The probability of finding sufficiently high
velocity fluctuations at this restricted flame surface area was
derived separately in the previous section. We define now
\begin{equation}
  \label{eq11}
  A_\mathrm{det}(t) = A_\mathrm{flame}^{*}(t) P[v'(\ell_\mathrm{crit}) \geq v'_\mathrm{crit}](t)
\end{equation}
as the part of the flame surface area that can potentially undergo a
DDT \citep[see also][]{roepke2007d}. This quantity has to exceed a
critical value $A_\mathrm{crit}$ that is required for a DDT. We assume
that a DDT region has a smooth two-dimensional geometry and use
therefore $A_\mathrm{crit} = \ell_\mathrm{crit}^2 =
\unit[10^{12}]{cm^2}$. For $A_\mathrm{det}(t) > A_\mathrm{crit}$, we
finally check whether this condition holds for at least
$\tau_\mathrm{eddy_{1/2}}(\ell_\mathrm{crit})$ to ensure a sufficient
mixing (see Section~\ref{Sect.2}). If this is true, our DDT criterion
is met and detonations are initialized.  The number of DDTs
${N}_\mathrm{DDT}$ in our model is given by
\begin{equation}
  \label{eq12}
  {N}_\mathrm{DDT} = \frac{A_\mathrm{det}(t)}{A_\mathrm{crit}} ,
\end{equation}
where ${N}_\mathrm{DDT}$ is always rounded down to the next
integer. We note that both quantities $A_\mathrm{flame}^{*}(t)$ and
particularly \mbox{$P[v'(\ell_\mathrm{crit}) \geq \unit[10^8]{cm\
    s}^{-1}](t)$} may rise steeply within
$\tau_\mathrm{eddy_{1/2}}(\ell_\mathrm{crit})$, hence we often get
${N}_\mathrm{DDT} > 1$. The minimum time between two DDTs is given by
$\tau_\mathrm{eddy_{1/2}}(\ell_\mathrm{crit})$, since, after a
successful DDT, the time for $A_\mathrm{det}(t) > A_\mathrm{crit}$ is
restarted. The same holds for the case $A_\mathrm{det}(t) <
A_\mathrm{crit}$ happens before
$\tau_\mathrm{eddy_{1/2}}(\ell_\mathrm{crit})$ is reached.

We still have to decide on the location where detonations are
initialized.  Since the high turbulent velocity fluctuations are
crucial for a DDT, we chose those ${N}_\mathrm{DDT}$ grid cells from
$N_\mathrm{flame}^{*}(t)$ that contain the highest values of
$v'(\ell_\mathrm{crit})$.  In analogy to the deflagration ignition,
detonations are set by initializing an additional level set that
propagates supersonically at the appropriate detonation speed
\citep[see][]{fink2010a} through the white dwarf matter.

A shortcoming of this DDT model is that it does not assess whether
there is indeed a ``connected'' region of size $\unit[10^{12}]{cm^2}$
that fulfills the requirements for a DDT. The probability
\mbox{$P[v'(\ell_\mathrm{crit}) \geq v'_\mathrm{crit}](t)$} and the
flame surface area $A_\mathrm{flame}^{*}(t)$ are determined from all
(possibly disconnected) grid cells suitable for a DDT. Therefore they
do not provide any information on localized areas. They rather are
global quantities.  The same holds for
$\tau_\mathrm{eddy_{1/2}}(\ell_\mathrm{crit})$, since here we also use
a uniform value.

From a computational point of view, we emphasize that the inclusion of
$\tau_\mathrm{eddy_{1/2}}(\ell_\mathrm{crit})$ is also important to
keep the DDT criterion independent of resolution. Since the maximum
time step $\Delta_\mathrm{CFL}$ of our code is given by the
Courant-Friedrichs-Lewy (CFL) condition \citep{courantfriedrichs}, the
time steps of higher resolved simulations are shorter than for lower
resolved ones. Applying our criterion without a time-dependent
variable would mean that higher resolved simulations get an enhanced
chance for a successful detonation, simply because it tests for DDTs
more frequently.  We note that in our simulations
$\Delta_\mathrm{CFL}$ is usually much shorter than
$\tau_\mathrm{eddy_{1/2}}(\ell_\mathrm{crit})$.

\section{The fractal dimension of the flame and resolution test in one
  full-star model}
\label{Sect.4}

\begin{figure*}
  \begin{center}
    \subfigure[Number of all grid cells at the flame front for two
    different resolutions (thick curves) as well as theoretical curves
    if a specific fractal dimension is assumed.]
    {\includegraphics[angle=270,width=\columnwidth]{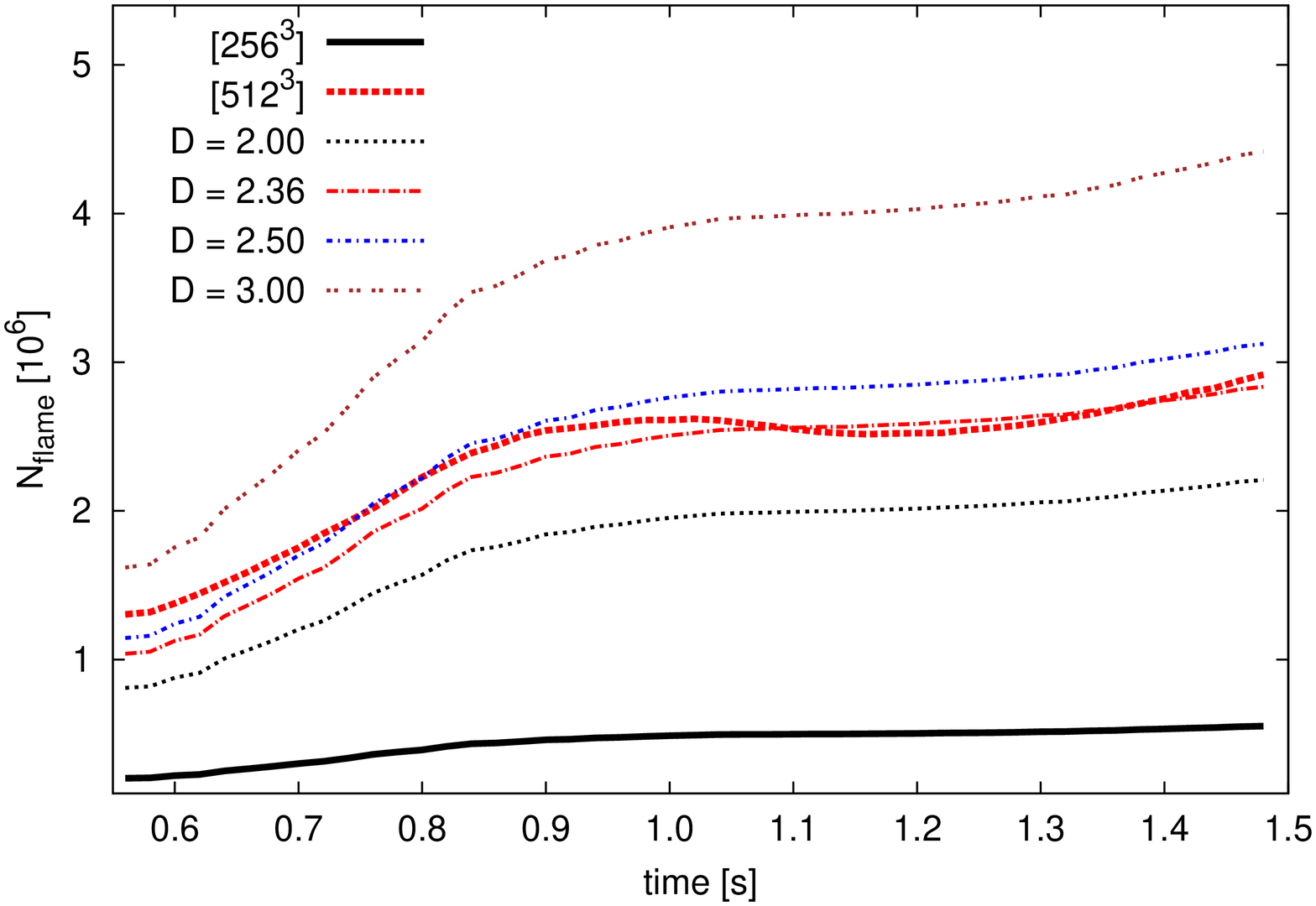}}
    \subfigure[Fractal dimension $D$ of the flame as function of
    time.]  {\includegraphics[angle=270,width=\columnwidth]{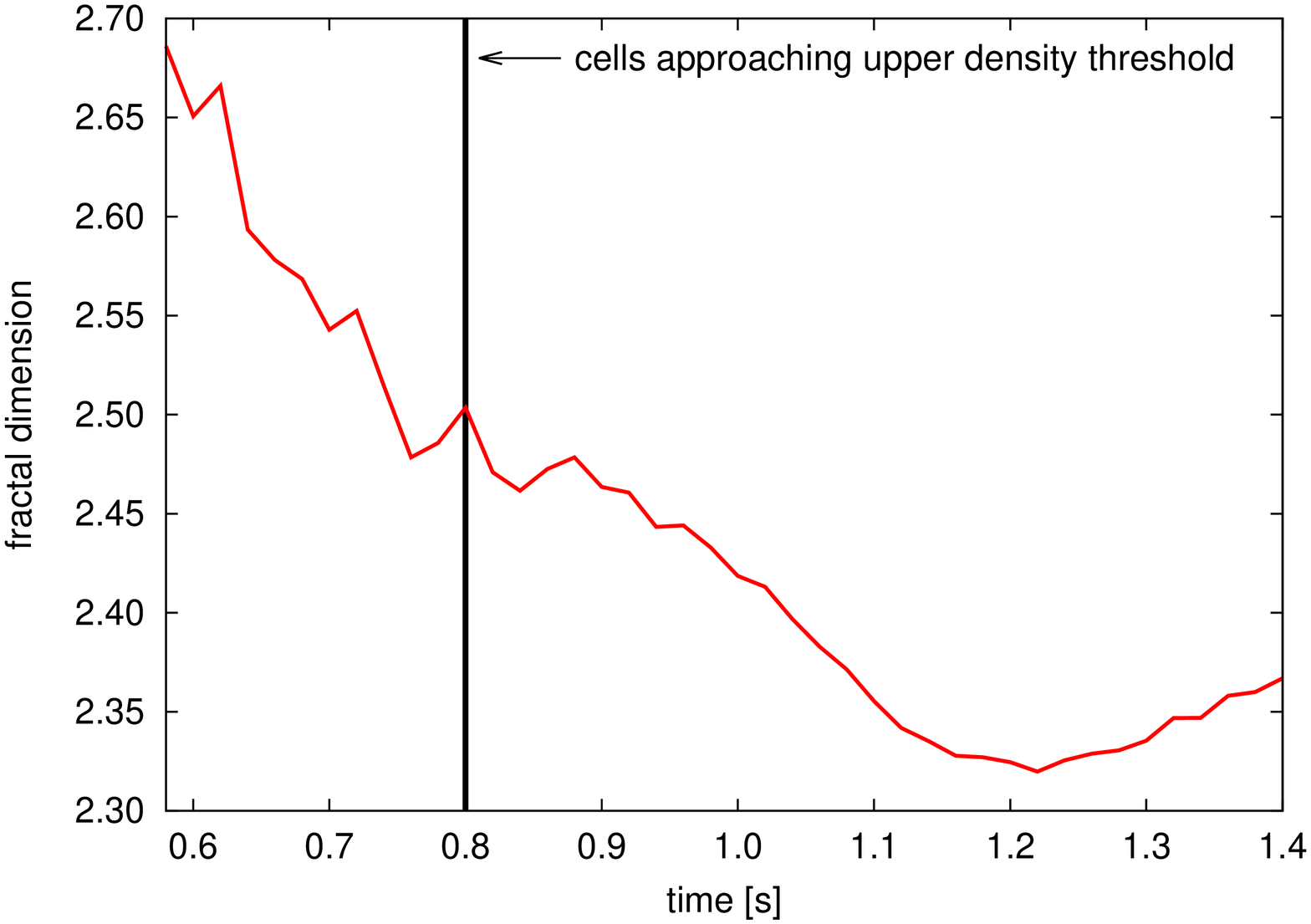}}
    \subfigure[Flame surface area $A_\mathrm{flame}^{*}(t)$ for
    different resolutions with $D=2.36$.]
    {\includegraphics[angle=270,width=\columnwidth]{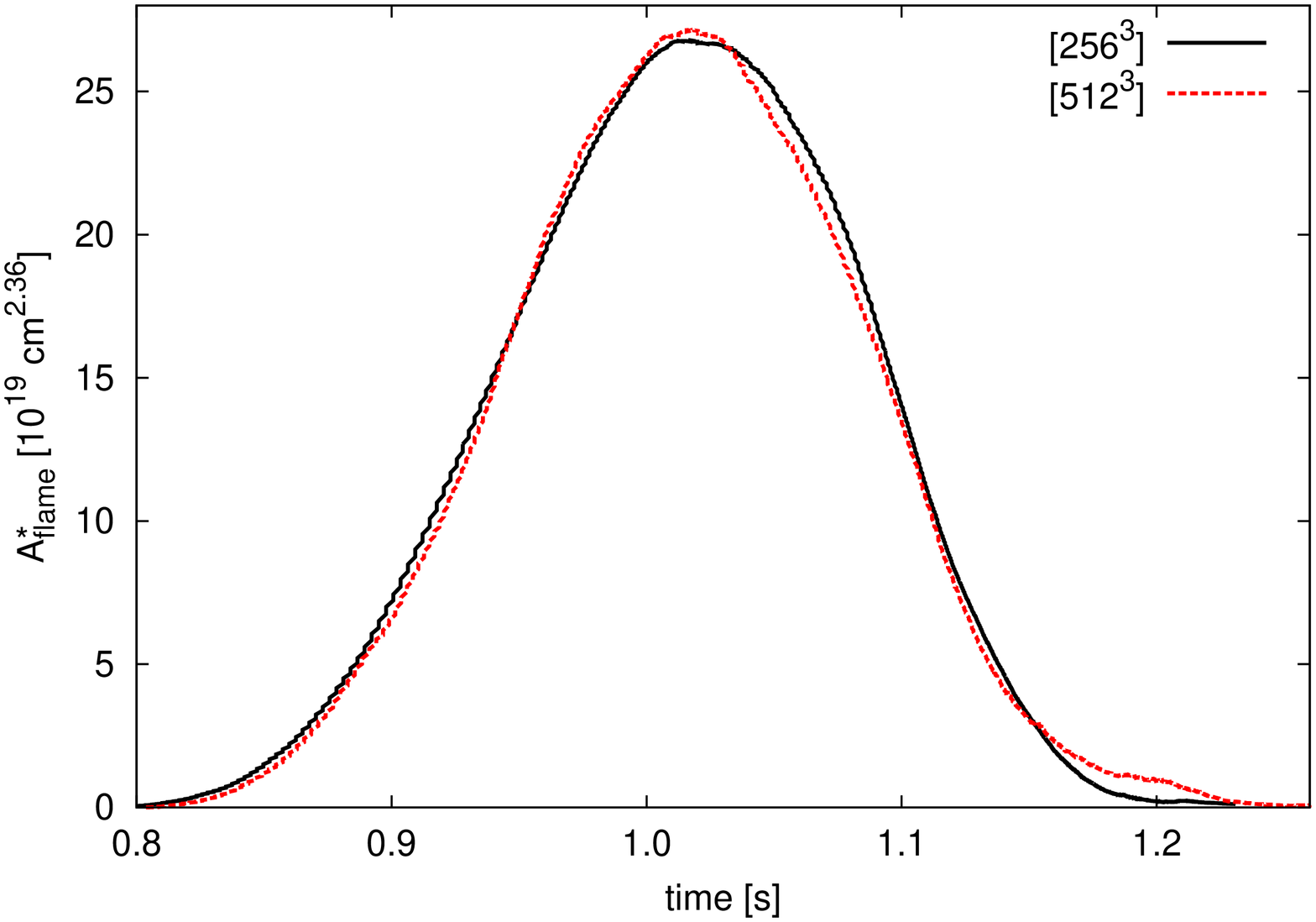}}
    \subfigure[Histograms of $v'(\ell_\mathrm{crit})$ at the flame
    front and the corresponding fits (Eq.~\ref{eq9}) for different
    resolutions.]
    {\includegraphics[angle=270,width=\columnwidth]{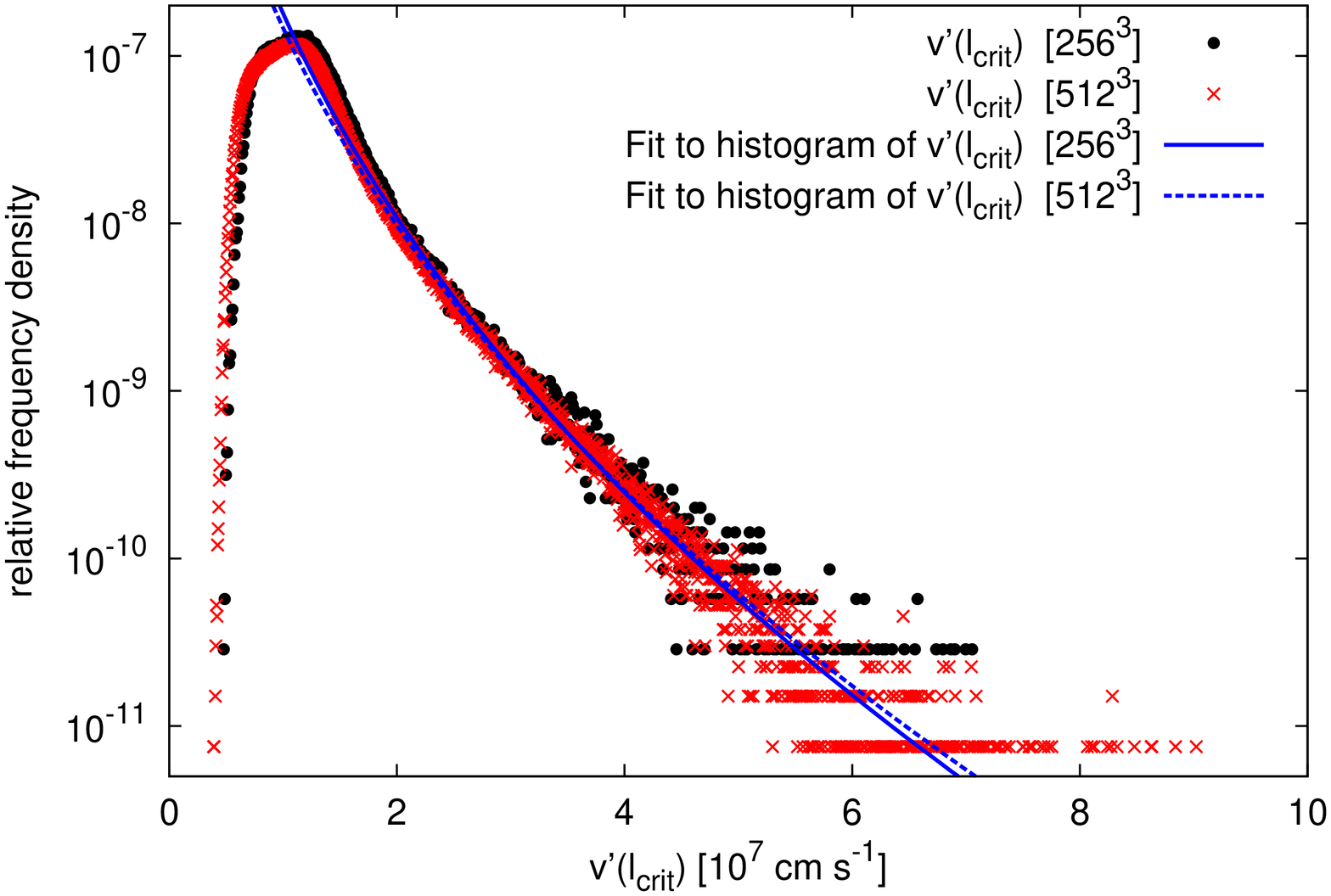}}
    \caption{Analysis of the fractal dimension of the flame and
      resolution dependence of the histogram of
      $v'(\ell_\mathrm{crit})$. (a) The number of grid cells found at
      the flame front indicates that the flame has a fractal
      character. (b) The fractal dimension is a time-dependent
      quantity which drops below 2.5 when the first grid cells
      approach the upper fuel density threshold of $\unit[1.5 \times 10^7]
      {g\ cm^{-3}}$. (c) The flame surface area
      $A_\mathrm{flame}^{*}(t)$ using a fractal dimension of 2.36 is
      very similar in both simulations. (d) The histograms and the
      corresponding fits of both simulations are in a good agreement.}
    \label{fig:2}
  \end{center}
\end{figure*}

\begin{figure*}
  \begin{center}
    \subfigure[] {\includegraphics[angle=270,width=\columnwidth]{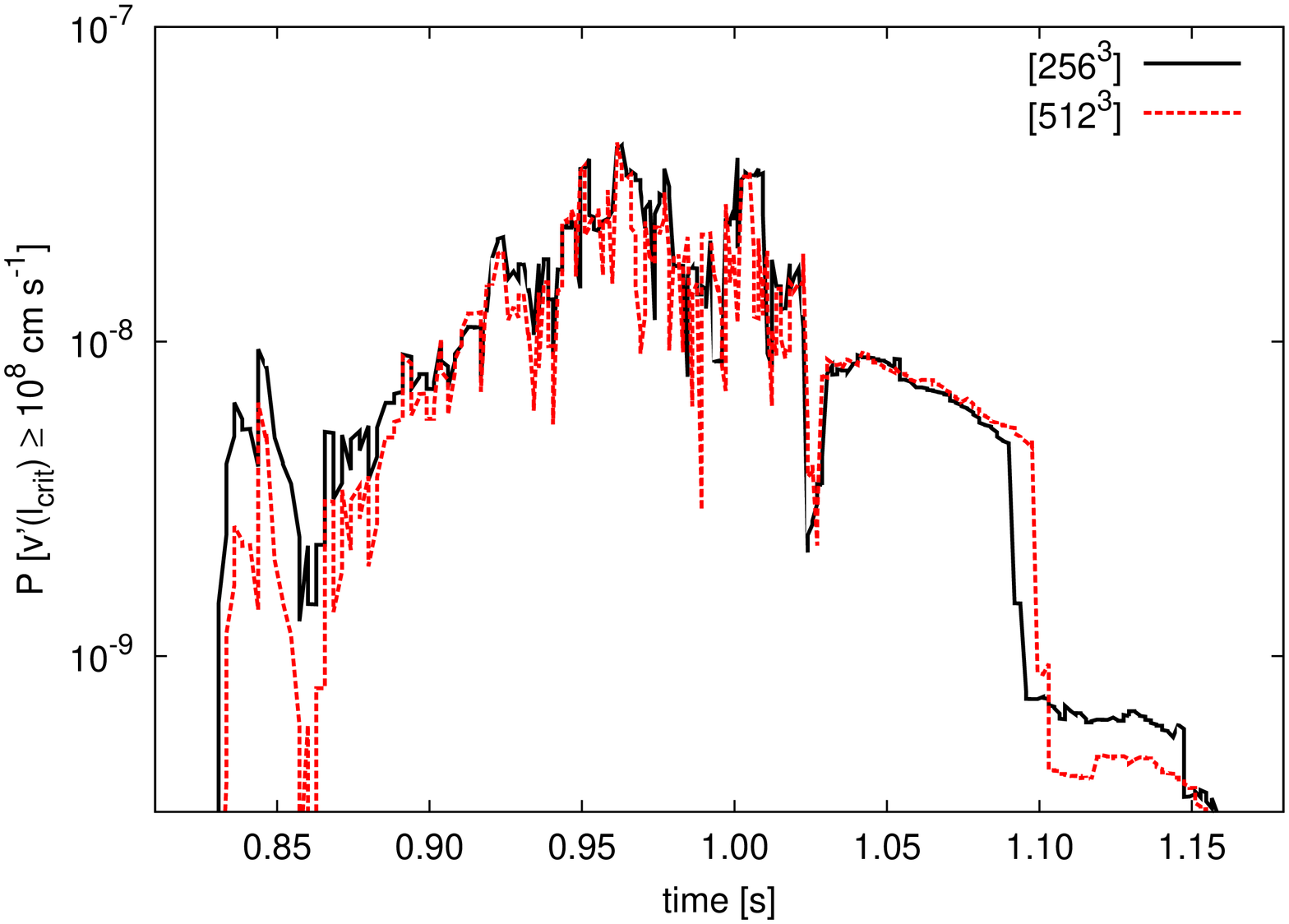}}
    \subfigure[] {\includegraphics[angle=270,width=\columnwidth]{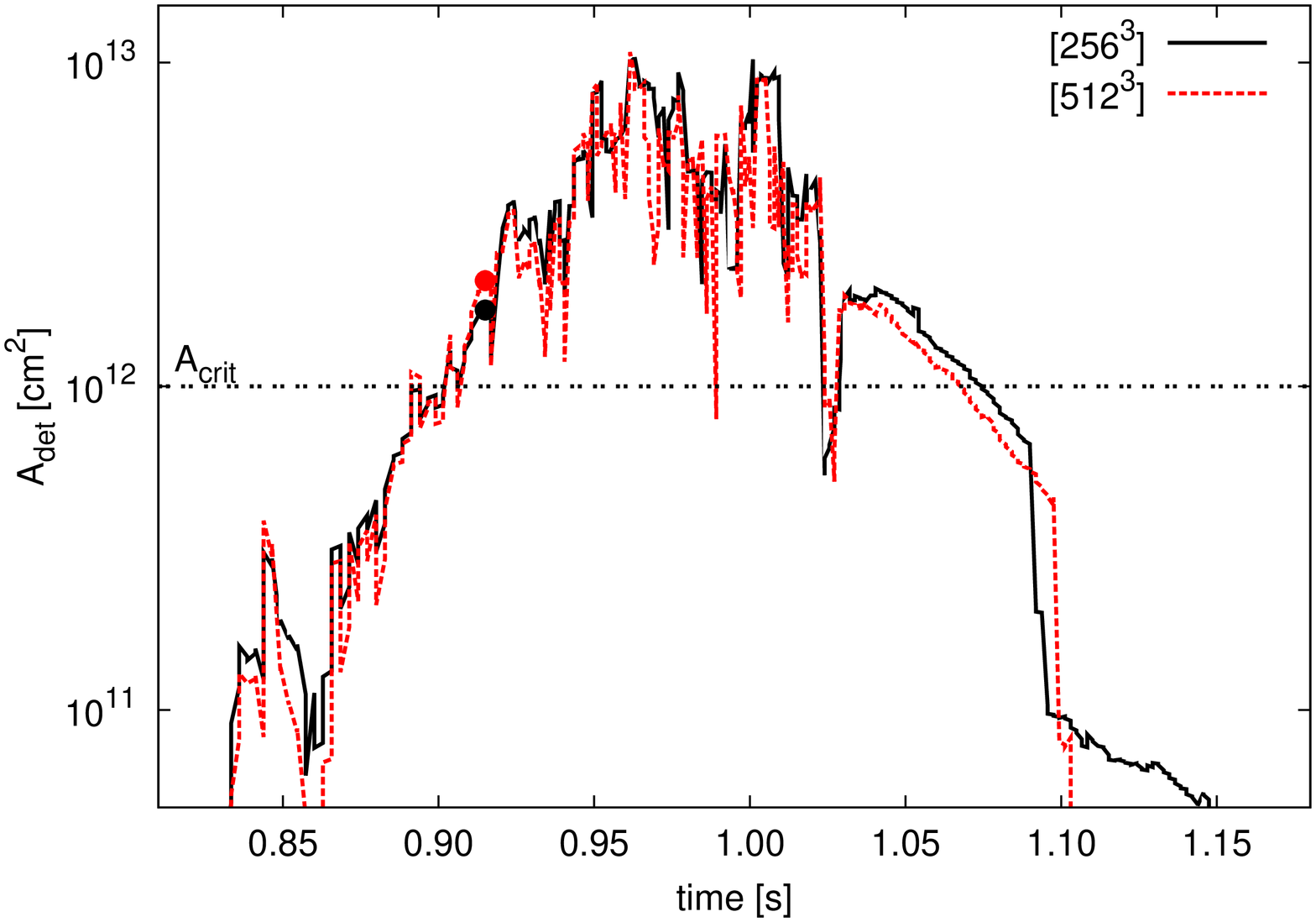}}
    \caption{(a) The probability $P[v'(\ell_\mathrm{crit}) \geq
      \unit[10^8]{cm\ s^{-1}}(t)$ of finding velocity fluctuations
      higher than $\unit[10^8]{cm\ s^{-1}}$ and (b) the size of the
      potential detonation area $A_\mathrm{det}(t)$. For most of the
      time between $t = \unit[0.90]{s}$ and $t = \unit[1.07]{s}$,
      $A_\mathrm{det}(t) > A_\mathrm{crit}$ holds. The DDT criterion
      is met for the first time at $t\approx\unit[0.92]{s}$ in both
      simulations (see dots at the curve of $A_\mathrm{det}(t)$).}
    \label{fig:3}
  \end{center}
\end{figure*}

To test the resolution dependence of the implemented DDT criterion we
apply it to the deflagration model described in Section~\ref{Sect.3.1}
and run it with a resolution of $256^3$ and $512^3$ grid
cells. Unfortunately, we cannot perform a detailed resolution study,
since simulations with more than $512^3$ grid cells are
computationally too expensive, while the DDT model cannot be applied
for very low resolved simulations due to insufficient data for fitting
the histogram of $v'(\ell_\mathrm{crit})$. The quantities and the
corresponding threshold values of the DDT criterion shall be
summarized here: $1/3 \leq X_\mathrm{fuel} \leq 2/3$, $0.5 \lesssim
\rho_\mathrm{fuel} / (10^7 \mathrm{g\ cm^{-3}}) \lesssim 1.5$,
$v'_\mathrm{crit} = \unit[10^8]{cm\ s^{-1}}$, $A_\mathrm{crit} =
\unit[10^{12}]{cm^2}$ and $\tau_\mathrm{eddy_{1/2}}(\ell_\mathrm{crit}) =
\unit[5 \times 10^{-3}]{s}$. One parameter still undetermined is the
fractal dimension of the flame, which we now derive from the
resolution test.

\subsection{The fractal dimension of the flame}
\label{Sect.4.1}
In Fig.~\ref{fig:2}(a) we show $N_\mathrm{flame}(t)$.  The thick black
curve is the result for the lower resolved simulation and the thick
red (dashed) curve for the higher resolved one, respectively. The
other curves are theoretically expected results for the higher
resolved simulation, if a certain fractal dimension of the flame is
assumed. These curves can be calculated from $N_\mathrm{flame_1}(t)$
and the known resolutions $\Delta_1(t)$ and $\Delta_2(t)$ of the
simulations, by specifying a value for $D$ in Eq.~\ref{eq4}. We see
that the curves for $D=2$ and $D=3$ are not consistent with the data,
which shows that the flame is indeed a fractal.

In Fig.~\ref{fig:2}(b) the fractal dimension $D$ (calculated from
Eq.~\ref{eq5}) is shown as function of time. A necessary constraint in
our criterion is that $\rho_\mathrm{fuel}$ must be in a certain range
(see Section~\ref{Sect.2}). At approximately $t = \unit[0.8]{s}$, the
first grid cells at the flame front approach
$\rho_\mathrm{fuel} = \unit[1.5\times 10^7]{g\ cm^{-3}}$,
while most part of the flame resides at
higher densities. We see that at this time $D\approx 2.5$, but DDTs
will occur later when a sufficiently large part of the flame surface
area meets the DDT constraints and
$\tau_\mathrm{eddy_{1/2}}(\ell_\mathrm{crit})$ has elapsed. At these
times $D < 2.5$. In agreement with theory \citep[e.g.][]{halsey1986a,
  sreenivasan1991a}, we use a constant value of $D = 2.36$ for our DDT
model (see Section~\ref{Sect.3.2}). In Fig.~\ref{fig:2}(c), we show
the quantity $A_\mathrm{flame}^{*}(t)$ (calculated from Eq.~\ref{eq3}
with $D = 2.36$) for both simulations.  Since the curves are in a good
agreement, the choice of $D = 2.36$ is justified.

We next discuss some caveats concerning the determination of the
fractal dimension of the flame. It is obviously a rough approximation
to take $D$ as a constant, since we see in Fig.~\ref{fig:2}(b) that
this quantity declines continuously until $t\approx 1.2$ s. Moreover,
one could expect that models with a different deflagration phase may
have another curve progression of $D$. Therefore, a determination of
$D$ for every delayed detonation simulation that has a different
evolution of the deflagration would be necessary. However, at the time
when DDTs occur, we assume that turbulence in the deflagration phase
is fully developed and obeys well defined statistical
properties. Hence a significant deviation of $D$ for different
deflagrations seems to be unlikely, but it cannot be ruled out
completely. Finally the results may be different, if higher
resolutions with more data are used for the determination of $D$, so
that a convergence of $D$ may be obtained at very high resolutions
only. Apart from the limited computational resources that prevent
simulations at such high resolutions, we mention in
Section~\ref{Sect.3.2} that the flame is no ideal fractal, so that at
some length scale Eq.~\ref{eq5} becomes inappropriate to derive
$D$. 

We note that the values for $D$ in
Fig.~\ref{fig:2}(b) may also indicate that the flame is affected by
different mechanisms and instabilities that drive the turbulence,
since $2.3 \lesssim D \lesssim 2.5$ are expected for different
instabilities at the flame front (see Section~\ref{Sect.3.2}).
In any case, the chosen value of $D = 2.36$ is appropriate for our
purposes, since both curves of $A_\mathrm{flame}^{*}(t)$ in
Fig.~\ref{fig:2}(c) are in a good agreement.

\subsection{The probability of finding high velocity fluctuations}
\label{Sect.4.2}
In Fig.~\ref{fig:2}(d), a histogram of $v'(\ell_\mathrm{crit})$ with
the fit according to Eq.~\ref{eq9} through the data for both
simulations is shown at $t = \unit[0.9]{s}$. Since we are mainly
interested in the high velocity fluctuations, the starting point of
the fit is at twice the velocity at the maximum of the corresponding
histogram. We see that the histograms and the approximated PDFs are in
very good agreement. For the lower resolved simulation, however, the
data contains more scatter and there is an earlier cutoff toward
higher velocity fluctuations. This is the result of a coarser binning
due to less data in lower resolved simulations.

The probability $P[v'(\ell_\mathrm{crit}) \geq \unit[10^8]{cm\
  s}^{-1}](t)$ is shown in Fig.~\ref{fig:3}(a) for both simulations.
The good agreement of both curves reflects that the approximated PDFs
are largely independent of resolution. The highest values are found
for $\unit[0.95]{s} < t < \unit[1.00]{s}$, when the intermittency in
the turbulence is most pronounced.

\subsection{The detonation area}
\label{Sect.4.3}
The quantity $A_\mathrm{det}(t)$ is shown in Fig.~\ref{fig:3}(b) for
both resolutions.  Since $A_\mathrm{det}(t)$ is calculated from
$A_\mathrm{flame}^{*}(t)$ and $P[v'(\ell_\mathrm{crit}) \geq
\unit[10^8]{cm\ s}^{-1}](t)$, it is also independent of resolution.
This manifests in Fig.~\ref{fig:3}, in which the overall shape of the
curve of $A_\mathrm{det}(t)$ is very similar to that of
$P[v'(\ell_\mathrm{crit}) \geq \unit[10^8]{cm\ s}^{-1}](t)$. In
particular the strong variations of $P[v'(\ell_\mathrm{crit}) \geq
\unit[10^8]{cm\ s}^{-1}](t)$ can be identified again. This indicates
that the change of the flame surface area $A_\mathrm{flame}^{*}(t)$
for a given interesting time interval
$\tau_\mathrm{eddy_{1/2}}(\ell_\mathrm{crit})$ is much smaller than
the fast temporal variations of the probability
$P[v'(\ell_\mathrm{crit}) \geq \unit[10^8]{cm\ s}^{-1}](t)$. To see
this, compare $A_\mathrm{flame}^{*}(t)$ in Fig.~\ref{fig:2}(c) with
Fig.~\ref{fig:3}(a).

For our resolution test we have chosen
$A_\mathrm{crit}=\unit[10^{12}]{cm^2}$, and we see that this value is
exceeded by $A_\mathrm{det}(t)$. Therefore, DDTs will occur if
$A_\mathrm{det}(t) > A_\mathrm{crit}$ for at least
$\tau_\mathrm{eddy_{1/2}}(\ell_\mathrm{crit})$. This condition is
indeed reached in both simulation, where the first DDTs are
initialized at approximately $\unit[0.92]{s}$. This time is marked
with a dot at the curve of $A_\mathrm{det}$ in Fig.~\ref{fig:3}(b).
For the lower resolved simulation we find $A_\mathrm{det}\approx 1.72
\times 10^{12}$, hence ${N}_\mathrm{DDT} = 1$. In the higher resolved
simulation we find $A_\mathrm{det}\approx 2.12 \times 10^{12}$, hence
here detonations are initialized already in two grid cells. Note that
these cells are located at different deflagration plumes and that they
are spatially disconnected.  As long as
$A_\mathrm{det}(t) > A_\mathrm{crit}$ for
$\tau_\mathrm{eddy_{1/2}}(\ell_\mathrm{crit})$, new DDTs commence at
later time steps. This happens in our simulations, since
$A_\mathrm{det}(t) > A_\mathrm{crit}$ for most of the time in the
interval $0.92 \mathrm{s} \lesssim t \lesssim 1.07
\mathrm{s}$. However, there are a few interruptions: In some time
steps the condition $A_\mathrm{det}(t) < A_\mathrm{crit}$ occurs
before $\tau_\mathrm{eddy_{1/2}}(\ell_\mathrm{crit})$ was reached,
preventing some DDTs. The maximum of ${N}_\mathrm{DDT}$ is 10 for the
lower resolved and 12 for the higher resolved simulation, which is
reached at $t \approx\unit[0.96]{s}$ for both cases. The last DDT
occurs at $t\approx\unit[1.06]{s}$ in a single grid cell in both
simulations.

\begin{figure*}
  \begin{center}
    \subfigure[Simulation with $256^3$ grid cells]
    {\includegraphics[width=\columnwidth]{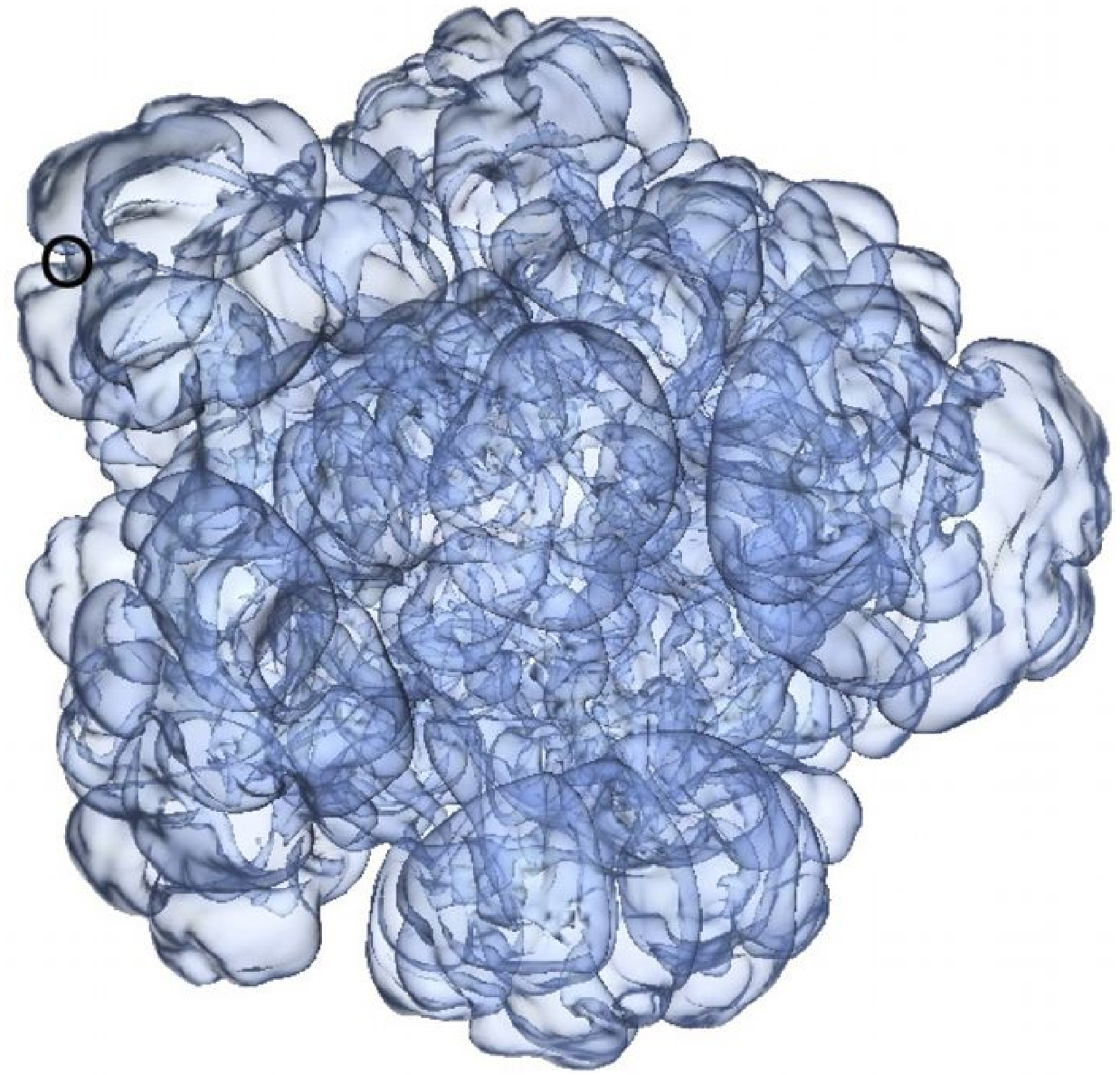}} \subfigure[Simulation
    with $512^3$ grid cells]
    {\includegraphics[width=\columnwidth]{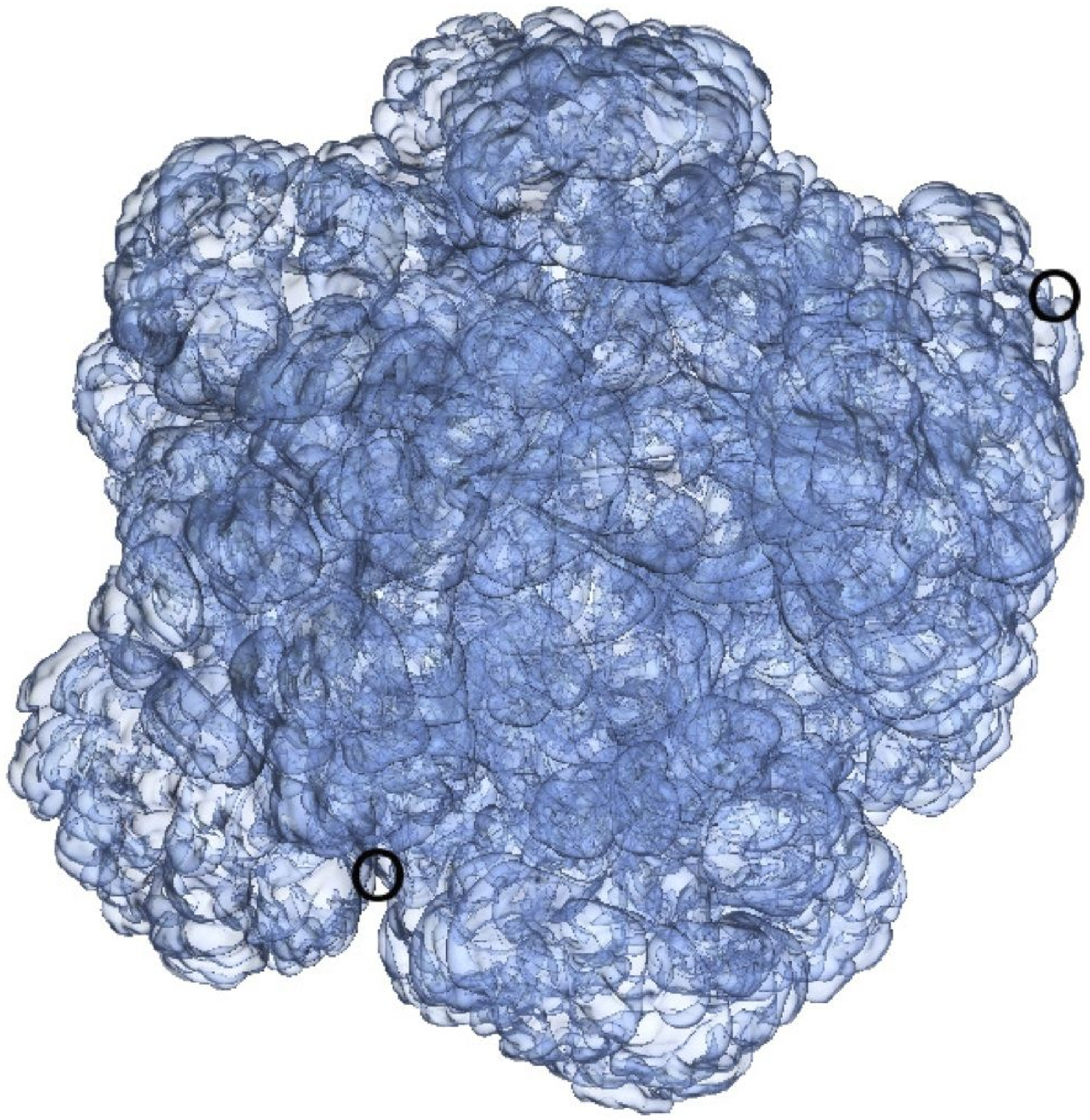}}
    \caption{Shown is the deflagration flame (transparent iso-surface)
      at the time of the first DDT for both simulations. The DDT spots
      are encircled.}
    \label{fig:visddt}
  \end{center}
\end{figure*}

In Fig.~\ref{fig:visddt}, the time of the first DDT is visualized for
both simulations. The deflagration flame, represented by the level set
function, is shown as a transparent iso-surface. While the number of
DDTs is largely resolution-independent, their localization is
generally quite different. In Fig.~\ref{fig:visddt}, the first DDT
occurs at different places at the deflagration flame (both figures
show the same viewing angle). The reason is that the exact number and
locations of the grid cells in which the highest velocity fluctuations
occur differs. In subsequent papers of this series we will show that
different distribution of DDT spots have an impact on the
\isotope[56]{Ni} production rate in the detonation phase.

\section{Conclusion}
\label{Sect.5}
We introduced the first subgrid-scale model for implementing
deflagration-to-detonation transitions (DDT) in a hydrodynamic code
for large-scale simulations of Type Ia supernova (SN~Ia)
explosions. The model includes the current knowledge on DDTs in SNe~Ia
and can be summarized as follows: We first ensure that a sufficient
number of grid cells at the flame have a certain fuel fraction and are
in a certain fuel density range. From the number and size of these
cells we determine a suitable flame surface area for DDTs, where we
assume that the flame can be considered as a fractal. Simultaneously,
we construct a histogram of the turbulent velocity fluctuations in the
above-mentioned cells, where we rescale these fluctuations from the
grid scale to the critical length scale of a DDT region by assuming
Kolmogorov turbulence. Then we estimate the probability of finding
sufficiently high velocity fluctuations for a DDT, by applying a fit
function to the histogram. This probability multiplied with the flame
surface area that is suitable for a DDT constitutes a potential
detonation area, which we compare with the required critical size of a
DDT region. When the potential detonation area exceeds this critical
size for at least half of an eddy turnover time, the DDT constraints
are fulfilled. In this case detonations are initialized in the grid
cells at the flame surface area suitable for a DDT that contain the
highest velocity fluctuations. The number of initialized detonations
equals the ratio of the potential detonation area to the critical size
of a DDT region.

Although our model refers to the initiation of the detonation via
  the Zel'dovich gradient mechanism, we note that other proposed
  mechanisms for forming a detonation out of a turbulent deflagration
  burning regime \citep{poludnenko2011a, kushnir2012a} would require a similar parameterization of the
  DDT-SGS model. In all cases, the critical quantity is the strengh of
  turbulence. However, 
  the models of \citet{poludnenko2011a} and
  \citet{kushnir2012a}  require turbulence speeds close to sonic, which we do not observe in our
  simulations of deflagrations in white dwarfs. The velocity
    fluctuations of $10^{8} \, \mathrm{cm}\,\mathrm{s}^{-1}$ assumed
  in our DDT-SGS model correspond to Mach numbers in the density range
  in which DDTs are expected of $\sim$$0.3$ with
  respect to the fuel material ($\sim$$0.1$ with respect to the ashes).

We showed that the DDT-SGS model is largely resolution
independent. Assuming that the DDT
region has a smooth two-dimensional geometry we found in a specific
deflagration model that the criterion is met, indicating that the
necessary constraints for DDTs in SNe~Ia were appropriate. Our model
includes a global criterion, since the histogram of
$v'(\ell_\mathrm{crit})$ and $A_\mathrm{flame}^{*}(t)$ do not provide
any information of local areas. Therefore, a shortcoming of our model
is that we cannot fully ensure that there is indeed a compact region
that obeys the necessary constraints for a DDT.

For testing our DDT model, we used one specific simulations of the
deflagration phase in a thermonuclear explosion of a
Chandrasekhar-mass WD. The evolution of the turbulent deflagration
depends strongly on the ignition scenario of the flame, which is
currently unknown. Certain turbulent deflagrations 
will meet a given DDT criterion more frequently, 
which will consequently affect the occurrence of DDTs. 
Therefore, the ignition scenario of the
deflagration is another crucial model parameter for 
simulations of delayed detonations. An analysis of 
the importance of the ignition scenario on the DDT SGS-model 
will be the subject of a future publication. 
The values of the DDT parameters are not well
known and have been kept constant or fixed in a certain range in our
DDT model. For this reason, we intend to perform a
parameter study by varying all DDT quantities. These future
studies will reveal further insights into the relevance and constraints
of delayed detonations in Chandrasekhar-mass white dwarfs.

\begin{acknowledgements}
This work was partially supported by 
the Deutsche Forschungs\-gemeinschaft via
the Transregional Collaborative Research Center TRR 33 ``The Dark
Universe'', the Excellence Cluster EXC153 ``Origin and Structure of
the Universe'', the Emmy Noether Program (RO 3676/1-1) and the graduate school ``Theoretical Astrophysics and Particle Physics'' GRK 1147.
FKR also acknowledges financial support by the ARCHES prize of the 
German Ministry of Education and Research (BMBF) and by
the Group of Eight/Deutscher Akademischer Austausch Dienst (Go8/DAAD)
German-Australian exchange program. 
\end{acknowledgements}

\end{document}